
\input phyzzx
\pubnum={KANAZAWA 93-04}
\date={March 1993}
\titlepage
\title{Equivalent models for gauged WZW theory
\footnote\dagger{Reseach supported in part by SASAKAWA SCIENTIFIC
RESEARCH GRANT of THE JAPAN SCIENCE SOCIETY}
}
\author{Haruhiko Terao
\footnote*{e-mail:terao@hep.s.kanazawa-u.ac.jp}
and Kiyonori Yamada
\footnote{**}{e-mail:yamada@hep.s.kanazawa-u.ac.jp}
}
\vskip 0.5cm
\address{Department of Physics, Kanazawa University
\break Kanazawa 920-11, Japan}

\abstract{
We modify the $SL(2,{\bf R})/U(1)$ WZW theory, which was shown to
describe strings in a 2D black hole, to be invariant under chiral
$U(1)$ gauge symmetry by introducing a Steukelberg field. We impose
several interesting gauge conditions for the chiral $U(1)$ symmetry.
In a paticular gauge the theory is found to be reduced to the
Liouville theory coupled to the $c=1$ matter perturbed by the
so-called black hole mass operator. Also we discuss the physical
states in the models briefly.
}

\endpage

\section{Introduction}
String theories have been investigated deeply as candidates for
the quantum gravity theory. Recently strings in two dimensional (2D)
target space have been attracting much attention and been studied
in various ways. Especially $SL(2,{\bf R})/U(1)$ WZW
theory, which was shown to describe strings on a 2D black hole
background by Witten,
\Ref\W{E. Witten, Phys. Rev. {\bf D44} (1991) 314.}
seems to give us a nice toy model to examine quantum effects to
dynamics of space-time such as space-time singularity, black hole
evapolation and so on.

In the model the 2D black hole space-time is sliced out from three
dimensional parameter space of $SL(2,{\bf R})$ group. The sliced
manifold depends on not only the choice of the variables but also
on the gauge conditions. Actually it has been pointed out that the
$SL(2,{\bf R})/U(1)$ model is related with the Liouville theory
with $c=1$ matter, which describes strings in a flat 2D space-time,
in the ref. [\W]. Thus the $SL(2,{\bf R})/U(1)$ WZW theory
may be expected to contain a family of string theories on various
background geometries.

The relations between the $SL(2,{\bf R})/U(1)$ theory and the
$c=1$ Liouville theory have been discussed by many other people.
\Ref\BK{M. Bershadsky and D. Kutasov, Phys. Lett. {\bf B266}
(1991) 345.}
\Ref\MS{E. J. Martinec and S. L. Shatashvili, Nucl. Phys.
{\bf B368} (1992) 338.}
\Ref\DVV{J. Dijkgraaf, H. Verlinde and E. Verlinde, Nucl. phys.
{\bf B371} (1992) 269.}
\Ref\DN{J. Distler and P. Nelson, Nucl. Phys. {\bf B374} (1992)
314.}
\Ref\EKY{T. Eguchi, H. Kanno and S. K. Yang, Phys. Lett. {\bf B298}
(1993) 73.}
\Ref\IK{H. Ishikawa and M. Kato, preprint UT-Komaba/92-11,
hepth-9211080.}
In the ref. [\DN] the WZW theory was found to contain the so called
new discrete states in the physical spectrum. On the other hand a
isomorphism in the algebraic structures, such as $W_{\infty}$ algebra,
the ground ring and BRST cohomology, between the two theories have
been found recently [\EKY,\IK]. Thus this problem has not been
completely cleared up yet.

In this note we also study the relations between the two models by
starting with the action of the gauged WZW theory. Originally in
the $SL(2,{\bf R})/U(1)$ WZW theory, the axial $U(1)$
symmetry is explicitly broken reflecting the axial anomaly. We
first restore the chiral $U(1)$ gauge symmetry by introducing a
Steukelberg field. Then it will be found that a paticular gauge
fixing reduces the WZW theory to the $c=1$ Liouville theory
perturbed by the so-called black hole mass operator. Also another
interesting gauge and the physical spectrum will be briefly examined.

\section{Chiral $U(1)$ gauged WZW theory}
The classical action of the $SL(2,{\bf R})/U(1)$ WZW theory
is given by \refmark\W
$$
\eqalign{
S_0 =&{k \over 8\pi} \int_{\Sigma}
      Tr[g^{-1}\partial_+ gg^{-1}\partial_- g]
      +{k \over 12\pi} \int_B Tr(g^{-1}dg)^3  \cr
     &-{k \over 4\pi} \int_{\Sigma}
      Tr[A_-g^{-1} \partial_+g +
         A_+ \partial_-gg^{-1} -
         A_+g A_-g^{-1} - 2A_+A_-], \cr
}
\eqno(1)
$$
where $g$ is a group element of $SL(2,{\bf R})$ and the gauge fields
$A_{\pm}$ belong to the $U(1)$ subalgebra generated by $\sigma_3$.
\footnote{1)}{
In this note we work on a Minkowski worldsheet and a Lorentz-signature
space-time. The Euclidean version will be obtained by an appropriate
continuation.
}
Here we parametrize the $SL(2,{\bf R})$ group element by using the
Gauss decomposition
$$
g= \left( \matrix{1&0 \cr
                  \chi&1 \cr} \right)\left( \matrix{ e^{\rho}&0 \cr
		  0&e^{-\rho} \cr} \right)\left( \matrix{1&\psi \cr
		  0&1 \cr} \right)
		  \eqno(2)
$$
with $\rho, \psi, \chi \in {\bf R}$. In this parametrization the
action (1) may be rewritten into \refmark\MS
$$
S_0 = {k \over 4\pi} \int d^2x
      [(\partial_+ \rho - A_+)(\partial_- \rho - A_-)
       +e^{2\rho}(\partial_+ + A_+) \chi (\partial_- + A_- ) \psi] .
\eqno(3)
$$
This action is invariant under a global
$SL(2,{\bf R}) \times SL(2,{\bf R})$
symmetry as well as a local (vector) $U(1)$ symmetry, whose
transformations are given by
$$
\eqalign{
\delta \rho    &= \epsilon , \cr
\delta \psi    &= -\epsilon\psi, \cr
\delta \chi    &= -\epsilon\chi \cr
\delta A_{\pm} &= \partial_{\pm}\epsilon . \cr
}
\eqno(4)
$$
If we impose the unitary gauge $\rho=0$ and integrate out the gauge
fields $A_{\pm}$, then the classical action turns out to be
$$
S = {-k \over 4\pi} \int d^2x
{\partial_+ u \partial_- v \over 1-uv }
\eqno(5)
$$
with $\chi=u$, $\psi=-v$.\refmark\MS
This is the string action on the Lorentzian black hole described
by the Kruskal coordinates $u$, $v$, which is given in the ref. [\W].

For the later purpose let us rewrite the action in first-order
formalism with respect to $\chi$ and $\psi$ by introducing canonical
momenta $\pi_{\chi}$ and $\pi_{\psi}$
$$
\eqalign{
S_0 ={1 \over 4\pi}\int d^2x
     &\biggl[k(\partial_+ \rho - A_+)(\partial_- \rho - A_-) \cr
     &+\pi_{\psi} (\partial_- + A_-) \psi
      +\pi_{\chi} (\partial_+ + A_+) \chi -
       {1 \over k}\pi_{\psi}\pi_{\chi}e^{-2\rho}\biggr]. \cr
}
\eqno(6)
$$
Now we may recover the broken axial $U(1)$ gauge symmetry by
introducing a Steukelberg field $\lambda$. The chiral $U(1)$
invariant action will be given by
$$
S = S_0 - {k \over 4\pi} \int d^2x
          (\partial_+ \lambda - A_+)(\partial_- \lambda - A_-).
\eqno(7)
$$
This action is indeed invariant under the following chiral gauge
transformations;
$$
\eqalign{
\delta\rho    &={1\over2}\theta_L + {1\over2}\theta_R ,   \cr
\delta\lambda &={1\over2}\theta_L + {1\over2}\theta_R ,   \cr
\delta\psi    &=-\theta_L \psi ,   \cr
\delta\chi    &=-\theta_R \chi ,   \cr
\delta A_{+}  &=\partial_+ \theta_L  ,   \cr
\delta A_{-}  &=\partial_- \theta_R .   \cr
}
\eqno(8)
$$

So far we have been discussing the classical action of the
$SL(2,{\bf R})/U(1)$ WZW theory. In the quantum theory we need
to take into account of quantum corrections to the classical action.
If we integrate out $\pi_{\psi}$, $\pi_{\chi}$, $\psi$ and $\chi$
in the action (7) then we will obtain the determinant factor,
$det[e^{-2\rho} (\partial_- - A_-)
e^{2\rho}(\partial_+ + A_+)]^{-1}$.
This determinant may be easily evaluated by using well-known
technique.
\Ref\GMOMS{A. Gerasimov, A. Morozov, M. Olshanetsky, A. Marshakov
and \nextline S. Shatashvili, Int. Journ. Mod. Phys. {\bf A5}
(1990) 2495.}
\refmark\MS
The $\rho$ dependent part of the determinant is found to be
$$
exp[{-1 \over 2\pi}\int d^2x
(\partial_+\rho \partial_-\rho
- A_+\partial_-\rho - A_-\partial_+\rho
-{1 \over 2}\sqrt{-\hat{g}}\hat{R}^{(2)}\rho )],
\eqno(9)
$$
where $\hat{g}$ denotes the reference metric in the conformal gauge.
On the other hand if we quantize the action (7) naively in the
operator formalism, then we will encounter "anomalies" in the
operator algebras of the $SL(2,{\bf R})$ currents and the energy-
momentum tensor. However when we add the quantum correction term
given by (9), then the currents will be found to satisfy proper
quantum algebras. Details in such analysis will be reported
elsewhere.
\Ref\TY{H. Terao and K. Yamada, in preparation.}

Anyhow the effective action in the conformal gauge would be given by
$$
\eqalign{
S = {1 \over 4\pi}\int d^2x
   &\biggl[(k-2)\partial_+ \rho \partial_- \rho
     +\sqrt{-\hat{g}}\hat{R}\rho
    -k\partial_+\lambda \partial_- \lambda \cr
   & +\pi_{\psi} \partial_- \psi +
     \pi_{\chi} \partial_+ \chi
     -{1\over k}\pi_{\psi} \pi_{\chi}e^{-2\rho}
     + b\partial_- c + \bar{b}\partial_+ \bar{c} \cr
   & + A_-\{\pi_{\psi}\psi -(k-2)\partial_+\rho
            +k\partial_+\lambda\} \cr
   & + A_+\{\pi_{\chi}\chi -(k-2)\partial_-\rho
            +k\partial_-\lambda\} \biggr], \cr
}
\eqno(10)
$$
where the ghosts $(c,b)$ are introduced to fix the diffeomorphism
and $\hbar$ is set to 1 simply. Moreover if we rescale the
variables as
$\rho = {1 \over \sqrt{2(k-2)}}\phi$,
$\lambda = {1 \over \sqrt{2k}}X$,
$\psi = a\gamma$,
$\pi_{\psi} = {1\over a}\beta$,
$\chi = b\bar{\gamma}$,
$\pi_{\chi} = {1 \over b}\bar{\beta}$,
this action (10) turns out to be
$$\eqalign{
S = {1\over 4\pi}\int d^2x
   &\biggl[{1\over2}\partial_+ \phi \partial_- \phi
     +{1 \over \sqrt{2(k-2)}}\sqrt{-\hat{g}}\hat{R}\phi
     -{1\over2}\partial_+ X \partial_- X  \cr
   & +\beta\partial_- \gamma + \bar{\beta}\partial_+ \bar{\gamma} +
     \mu \beta\bar{\beta}e^{-\sqrt{2\over{k-2}}\phi}
     +b\partial_- c + \bar{b}\partial_+ \bar{c} \cr
   & + A_-J_+ + A_+J_- \biggr],  \cr
}
\eqno(11)
$$
where $J_{\pm}$ are the chiral $U(1)$ gauge currents defined by
$$
\eqalign{
J_+ &= \beta\gamma - \sqrt{k-2 \over 2} \partial_+ \phi
       + \sqrt{k \over 2}\partial_+ X ,  \cr
J_- &= \bar{\beta}\bar{\gamma}
       - \sqrt{k-2 \over 2} \partial_- \phi
       + \sqrt{k \over 2} \partial_- X . \cr
}
\eqno(12)
$$
This action may be regarded as the chiral $U(1)$ gauged version of
the WZW action given in the ref. [\BK]. The parameter $\mu(\not=0)$,
which relates to the black hole mass, is completely arbitrary,
since it can be changed by the scaling. It should be noted that
the action (11) is not invariant under the chiral $U(1)$ transformations
classically, however it is invariant in the quantum level.

\section{Gauge fixing and $c=1$ string}
In this section we are going to perform gauge fixing to the chiral
$U(1)$ gauge symmetry in the quantum action (11) by using several
gauge conditions. First let us impose conditions $A_{\pm}=0$.
Through the standard BRST gauge fixing procedure we may find the
gauge fixed action to be
$$
\eqalign{
S = {1\over 4\pi}\int d^2x
    &\biggl[{1\over2}\partial_+ \phi \partial_- \phi
     +{1 \over \sqrt{2k'}}\sqrt{-\hat{g}}\hat{R}\phi
     -{1\over2}\partial_+ X \partial_- X   \cr
    & +\beta\partial_- \gamma + \bar{\beta}\partial_+ \bar{\gamma}
      +\mu \beta\bar{\beta}e^{-\sqrt{2\over{k'}}\phi} \cr
    & + b\partial_- c + \bar{b}\partial_+ \bar{c}
     + \eta\partial_-\xi + \bar{\eta}\partial_+\bar{\xi}
    \biggr],  \cr
}
\eqno(13)
$$
where $(\xi, \eta)$ are the ghost fields corresponding to the chiral
$U(1)$ parameter appeared in (8) and $k'$ denotes $k-2$ hereafter.
The BRST charge is also obtained by considering the chiral $U(1)$
transformations as well as diffeomorphism, and is found to be
\footnote{2)}
{Hereafter we are going to show only the left moving part of the
BRST charge, since the right moving part can be always given
in a similar form.}
$$
Q_{BRST} = \int d\sigma \{
           c T^{matter}_{++} + \xi J_+
           + cb\partial_+c + c\eta\partial_+\xi \} .
\eqno(14)
$$
Here $T^{matter}_{++}$ is the energy-momentum tensor in the matter
sector, which are given explicitly by
$$
\eqalign{
T^{matter}_{++} &= T^{c=1}_{++} + T^{\beta, \gamma}_{++}  \cr
                &= -{1 \over 2}(\partial_+\phi)^2
                   -{1 \over \sqrt{2k'}}\partial_+^2\phi
                   +{1 \over 2}(\partial_+X)^2
                   -\beta\partial_+\gamma.  \cr
}
\eqno(15)
$$
If we discard the term
${1 \over \mu}\beta \bar{\beta} exp(-\sqrt{{2 \over k'}}\phi)$
in the action (11), then this model turns out to be a free conformal
field theory. In this case the BRST charge (13) is exactly same as one
given in the ref. [\EKY,\IK]. The BRST nilpotency can be achieved
when $k=9/4$. However the effects of the discarded term can be
incorpolated by acting the screening operator
$\beta exp(-\sqrt{2 \over k'})\phi$,
which plays an important role to determine the physical spectrum and
to compute the correlation functions.
\Ref\DF{V. Dostenko and V. Fadeev, Nucl. Phys. {\bf B240} [FS12]
(1984) 312.}
\Ref\BF{D. Bernard and G. Felder, Commun. Math. Phys. {\bf 127}
(1990) 145.}

One of the gauge conditions which we would like to examine newly in
this note is
$$
\gamma = \bar{\gamma} =1.
\eqno(16)
$$
In this case the gauge fixed action is given by
$$
\eqalign{
S = {1\over 4\pi}\int d^2x
   &\biggl[{1\over2}\partial_+ \phi \partial_- \phi
     +{1 \over \sqrt{2k'}}\sqrt{-\hat{g}}\hat{R}\phi
     -{1\over2}\partial_+ X \partial_- X
     +\mu \beta\bar{\beta}e^{-\sqrt{2 \over k'}\phi} \cr
   & +A_- \{\beta - \sqrt{k' \over 2} \partial_+ \phi
            + \sqrt{k \over 2}\partial_+ X \}
     +A_+ \{\bar{\beta} - \sqrt{k' \over 2} \partial_- \phi
            + \sqrt{k \over 2}\partial_- X \} \cr
   & + b\partial_- c + \bar{b}\partial_+ \bar{c}
     + \eta\xi + \bar{\eta}\bar{\xi}
   \biggr].  \cr
}
\eqno(17)
$$
However here we may drop out the variables $\beta$, $\bar{\beta}$,
$A_{\pm}$, $\xi$, $\bar{\xi}$, $\eta$, $\bar{\eta}$ by using the
non-dynamical equations of motion,
$\beta       = \sqrt{{k' \over 2}}\partial_+ \phi
              - \sqrt{{k \over 2}}\partial_+ X, $
$\bar{\beta} = \sqrt{{k' \over 2}}\partial_- \phi
              - \sqrt{{k \over 2}}\partial_- X, $
$\eta = \bar{\eta} =0$,
$\xi = \bar{\xi} =0$.
After performing this the action turns out to be
$$
\eqalign{
S = {1\over 4\pi}\int d^2x
   &\biggl[{1\over2}\partial_+ \phi \partial_- \phi
     +{1 \over \sqrt{2k'}}\sqrt{-\hat{g}}\hat{R}\phi
     -{1\over2}\partial_+ X \partial_- X  \cr
   & +\mu \bigl(\sqrt{{k'}\over 2} \partial_+
         \phi - \sqrt{k \over 2}\partial_+ X \bigr)
         \bigl(\sqrt{{k'} \over 2} \partial_-
         \phi - \sqrt{k \over 2} \partial_- X \bigr)
         e^{-\sqrt{2\over{k'}}\phi} \cr
   & +b\partial_-c + \bar{b}\partial_+\bar{c} \biggr]. \cr
}
\eqno(18)
$$
Also we will find the BRST charge to be
$$
Q_{BRST} = \int d\sigma \{ cT^{c=1}_{++} + cb\partial_+c \},
\eqno(19)
$$
which is identical to the BRST charge of the Liouville theory
coupled to a $c=1$ matter $X$. The term like a gravitational vertex
operator in the action (18) is an $(1,1)$ operator called the
"black hole mass operator". \refmark\BK Thus we have seen that
the $SL(2,{\bf R})/U(1)$ WZW theory is completely equivalent to
the $c=1$ string theory perturbed by the black hole mass operator.
In this model also the screening operator
$\oint (\sqrt{{k' \over 2}}\partial \phi
- \sqrt{{k \over 2}}\partial X )
exp(-\sqrt{{2 \over k'}}\phi)$
coming from the black hole mass operator may be important to
calculate the correlation functions.

We may choose another gauge condition
$$
\beta = \bar{\beta} =1.
\eqno(20)
$$
By repeating similar manipulation the gauge fixed action and the BRST
charge are found to be
\footnote{3)}
{The Coulomb term would be deformed correspondingly
to the modification of the energy-momentum tensor.\refmark\TY}
$$
\eqalign{
S = {1 \over 4\pi}\int d^2x
   &\biggl[{1\over2}\partial_+ \phi \partial_- \phi
     +\bigl({1 \over \sqrt{2k'}} + \sqrt{k' \over 2}\bigr)
      \sqrt{-\hat{g}}\hat{R}\phi  \cr
   & -{1\over2}\partial_+ X \partial_- X
     -\sqrt{{k \over 2}}\sqrt{-\hat{g}}\hat{R}X  \cr
   & + b\partial_- c + \bar{b}\partial_+ \bar{c}
     +\mu e^{-\sqrt{2\over{k'}}\phi} \biggr] \cr
}
\eqno(21)
$$
and
$$
\eqalign{
Q_{BRST} = \int d\sigma \biggl[
          &c\biggl( -{1 \over 2}(\partial_+\phi)^2
                   -{1 \over \sqrt{2k'}}\partial_+^2\phi
                   +{1 \over 2}(\partial_+X)^2
                   -\sqrt{{k' \over 2}}\partial_+^2\phi
                   +\sqrt{{k \over 2}}\partial_+^2X
            \biggr) \cr
          & +cb\partial_+c  \biggr].  \cr
}
\eqno(22)
$$
This BRST charge also satisfies it's nilpotency if $k=9/4$.
At a glance this model looks to differ from the $c=1$ string. However
if we perform the following "Lorentz" transformation to the 2D target
space coordinate $(\phi,X)$,
$$
\eqalign{
\phi &= (k-1)\tilde{\phi} \pm \sqrt{k(k-2)}\tilde{X}, \cr
X    &= -\sqrt{k(k-2)}\tilde{\phi} \mp (k-1)\tilde{X}, \cr
}
\eqno(23)
$$
then the BRST charge (22) as well as the gauge fixed action (21)
written in terms of $\tilde{\phi}$ and $\tilde{X}$
are found to be identical to those of the $c=1$ string. At the
same time the "cosmological constant" term
$exp(-\sqrt{{2 \over k'}}\phi)$ is also transformed into an $(1,1)$
operator,
$V_{1-k,-k} = exp(\sqrt{{2 \over k'}}(1-k)\tilde{\phi} +
                 \sqrt{{2 \over k}}k\tilde{X})$.
Thus it is shown that the gauge $\beta = \bar{\beta} =1$ also
gives rise to the Liouville theory with $c=1$ matter perturbed
by the operator $V_{1-k,-k}$.

Ishikawa and Kato [\IK] pointed out that the BRST cohomology of the
$SL(2,{\bf R})$ $/ U(1)$ WZW theory is splitted into a product of BRST
cohomologies of the $c=1$ string and of a $U(1)/U(1)$ topological
theory. Our gauge choices kill out the topological freedom explicitly.

\section{The physical states}
If we admit naively that the string physics described by the gauged
WZW model does not depend on the choice of the gauge fixing conditions,
up to zero mode, then the three models found in the previous section
would be equivalent mutually. However it seems to be still necessary
and interesting to find out the explicit relations between these models
in order to see how strings in a black hole and strings in a flat
space-time are linked to each other.

In the $A_{\pm}=0$ gauge, the series of the physical states including
the so-called new discrete states have been found by Distler and
Nelson [\DN]. The simplest states among them are given in the form
of the $SL(2,{\bf R})$ primaries dressed by the Steukelberg
field $X$ (we consider only the "holomorphic" part below) as
\refmark\BK
$$
V_{j,m} = \gamma^r
           exp \biggl(\sqrt{{2 \over k'}}j\phi
                     -\sqrt{{2 \over k}}mX \biggr).
\eqno(24)
$$
The physical condition $Q_{BRST}\ket{phys} =0$ restricts the parameters
appearing in (24) to be $r=j-m$, $j=\pm{m\over3}-{1\over2}$ in the case
of $k=9/4$. We may obtain a series of discrete states ${\cal C}$ and
${\cal D}$ in the terminology in the ref. [\DN] by acting a operator
$$
I^- = \oint \gamma^{{3 \over 2}}e^{\sqrt{2}X}
\eqno(25)
$$
successively to the physical states given by $V_{j,m}$.

Another series of the discrete states can be obtained from
\refmark\BK
$$
\tilde{V}_{j,m} = \beta^s
                   exp \biggl(\sqrt{{2 \over k'}}j\phi
                             -\sqrt{{2 \over k}}mX \biggr),
\eqno(26)
$$
where $s$ is a positive integer and $s=-j+m$. For $k=9/4$ there are
two series of solutions of the physical state condition;
(a) $j={m \over 3} - {1 \over 4}$ and (b) $j=-{m \over 3} -1$.
The series (a) give a subset of the new discrete states in
$\tilde{\cal D}$
\refmark\DN. On the other hand the series (b) are contained in the
series ${\cal C}$ or ${\cal D}$.

We can show that these physical states are reduced to the tachyon
and the discrete states which appear in the $c=1$ string theory in
the case of other gauge fixing examined in the previous section.

It is obvious that $\gamma=1$ gauge reduces the states
$V_{j,m}$ in (24) and the series of the discrete states obtained
by acting the operator (25) on them to the tachyonic states and
the discrete states in the $c=1$ string theory. On the other hand
by substituing the equation of motion
$\beta = \sqrt{{k' \over 2}}\partial_+ \phi
        -\sqrt{{k \over 2}}\partial_+ X $,
the operators $\tilde{V}_{j,m}$ given in (26) are replaced by
$$
\tilde{V}_{j,m}=
\biggl({\sqrt2 \over4}\partial_+{\phi}
       -{3\sqrt2 \over4}\partial_+X \biggr)^s
exp \biggl(2\sqrt2 j\phi -{2\sqrt2 \over3}mX \biggr).
\eqno(27)
$$
We would like to examine only the lowest case $s=1$ here. The first
discrete state in the series (a) is given by $(j,m) = (1/8,9/8)$.
Then the operator $\tilde{V}_{{1 \over 8},{9 \over 8}}$ is found to
be a total divergence,
$$
\eqalign{
\tilde{V}_{{1\over8},{9\over8}}
&=\biggl({\sqrt2\over4}\partial_+{\phi}
         -{3\sqrt2 \over4}\partial_+X \biggr)
  exp \biggl({\sqrt2\over4}\phi - {3\sqrt2 \over4}X \biggr) \cr
&=\partial_+{\biggl(exp \biggl({\sqrt2 \over4}\phi
         -{3\sqrt2 \over4}X \biggr)\biggr)}, \cr
}
\eqno(28)
$$
which means the state given by $\tilde{V}_{{1\over8},{9\over8}}$
is BRST trivial. Also the first state in the series (b), $(j,m) =
(-1,0)$, is found to be the discrete state $W^{(-)}_{1,0}$
in the $c=1$ string theory up to a total divergence. Because
$$
\eqalign{
\tilde{V}_{-1,0}
&=\biggl({\sqrt2\over4}\partial_+{\phi}
         -{3\sqrt2\over4}\partial_+X \biggr)
  exp (-2\sqrt2 \phi) \cr
&=-{1\over8}\partial_+ exp(-2\sqrt2 \phi)
  -{3\sqrt2\over4}\partial_+X exp(-2\sqrt2 \phi).\cr
}
\eqno(29)
$$
Thus it is seen that the physical spectrum in the $SL(2,{\bf R})/
U(1)$ WZW theory indeed coincides with that in the $c=1$ string
theory in this level. At higher level $(s>1)$ we may expect a
similar mechanism to work to reduce the discrete states generated by
$\tilde{V}_{j,m}$ to BRST trivial states or to the physical states
of the $c=1$ string.

Next we shall examine the case of the $\beta=1$ gauge also.
After performing the "Lorentz" transformation $(23)$, the operators
given in (26) turn out to be
$$
\tilde{V}_{j,m}=
exp \biggl(2 \sqrt2 \tilde{j} \tilde{\phi}
          -{{2 \sqrt2} \over 3}\tilde{m} \tilde{X} \biggr),
\eqno(30)
$$
where
$\tilde{j}={5 \over 4}j - {1 \over 4}m$ and
$\tilde{m}=\pm ({1\over4}j - {5 \over 4}m )$.
{}From the relations of $j$ and $m$ for the states $\tilde{V}_{j,m}$
in $(26)$,
\footnote{4)}
{In this reduction we should suppose that the original
representation is given by a continuous parameter $s$ and
positive integers $r$.}
we will find $\tilde{j}=\pm{1 \over 3}\tilde{m} - {1 \over 2}$,
which are the condition that the operators have conformal dimension
one. Thus the states given by $(30)$ correspond to the tachyonic
states in the case of the $c=1$ string theory.

How about the states given by $V_{j,m}$ (24) in this gauge?
By using the equation of motion
$\gamma=\sqrt{{k' \over 2}} \partial_+ \phi
       -\sqrt{{k \over 2}}\partial_+ X $,
the operators $V_{j,m}$ may be represented as
$$
\biggl(-{{\sqrt2}\over4}\partial_+{\tilde{\phi}}
       \pm{3\sqrt2 \over4}\partial_+{\tilde{X}} \biggr)^r
exp \biggl({2\sqrt2}\tilde{j}\tilde{\phi}
          -{2\sqrt2 \over3}\tilde{m}\tilde{X}\biggr)
\eqno(31)
$$
through the Lorentz transformation. In the case of $r=1$, the
physical states are given by
$(j,m)= (-5/4,-9/4)$ and
$(j,m)= (-1/8,-9/8)$,
or equivalently
$(\tilde{j},\tilde{m}) =(-1,0)$ and
$(\tilde{j},\tilde{m})= (1/8,\pm 9/8)$.
Then the operators $V_{j,m}(r=1)$ are found to be reduced to
$$
\eqalign{
V_{-{5\over4},-{9\over4}} &= \pm{3\sqrt2 \over4}
\partial_+{\tilde{X}}exp(-{2\sqrt2}\tilde{\phi}), \cr
V_{-{1\over8},-{9\over8}} &= 0 \cr
}
\eqno(32)
$$
up to total divergences. The operator appearing in the right hand
side of (32) is just the $W^{(-)}_{1,0}$ written in terms of
$\tilde{\phi}$ and $\tilde{X}$.

Thus it seems to be true that the gauged WZW theory is completely
equivalent to the Liouville theory coupled to $c=1$ matter as
far as it is described in terms of the variables defined by
the Gauss decomposition (2). However the analysis of the
physical spectrum given here is far from complete. It would
be necessary to find out a complete description of the BRST
cohomology in terms of the free fields in order to confirm the
identical structure of the physical spectrum. Also the
implications from this observation to the 2D black hole physics
should be investigated.

We thank professor J. Kubo for valuable discussions.
\refout
\bye